\documentstyle[psfig]{aa}
\newcommand{\etal}{et al. }

\begin{document}
\title{A morphological
comparison between the central region in AGN and normal galaxies using HST
data\thanks{Based on observations made with the NASA/ESA Hubble Space
Telescope, obtained from the data archive at the Space Telescope
Science Institute. STScI is operated by the Association of Universities
for Research in Astronomy, Inc., under the NASA contract NAS5--26555.}}
\author{E.M. Xilouris\inst{1,2} \and I.E. Papadakis\inst{1,3} } 
\offprints{E.M. Xilouris, \email{xilouris@cea.fr}}
\institute{IESL, Foundation for Research and Technology--Hellas, P.O.Box
1527, 711 10 Heraklion, Crete, Greece
\and  DAPNIA/Service d' Astrophysique, CEA/Saclay, 91191 Gif--sur--Yvette
Cedex, France
\and Physics Department, University of Crete, 710 03 Heraklion, Crete, Greece}

\date{Received 30 January 2002 / Accepted 20 March 2002} 

\abstract{
We study the morphology of the central region of a sample of Active
Galactic Nuclei (AGN) and a ``control'' sample of normal galaxies using
archival observations of the WFPC2 instrument onboard the {\it Hubble
Space Telescope} (HST). We use the ellipse fitting technique in order to
get a good description of the inner ``smooth'' light distribution of the
galaxy. We then divide the observed galaxy image by the artificial image
from the fitted ellipses in order to detect morphological signatures in
the central region around the nucleus of the galaxy. We perform
quantitative comparisons of different subgroups of our sample of galaxies
(according to the Hubble type and the nuclear activity of the galaxies) by
calculating the average amplitude of the structures that are revealed with
the ellipse fitting technique. Our main conclusions are as follows:
\\ 
1) All AGNs show significant structure in their inner 100 pc and 1 kpc
regions whose amplitude is similar in all of them, independent of the
Hubble type of the host galaxy.
\\ 
2) When considering early-type galaxies, non-AGN galaxies show no
structure at all, contrary to what we find for AGN. 
\\ 
3) When considering late-type galaxies, both AGN and non-AGN galaxies show
significant structure in their central region.
\\
Our results are consistent with the hypothesis that all early-type
galaxies host a supermassive black hole, but only those that have enough
material in the central regions to fuel it show an active nucleus. The
situation is more complicated in late-type galaxies. Either not all of
them host a central black hole, or, in some of them, the material
inside the innermost 100 pc region is not transported to the scales of the
central engine for some reason, or the large amount of gas and dust hides
the active nucleus from our sight.
}
\titlerunning{Morphology of AGN and normal galaxies}
\authorrunning{Xilouris \& Papadakis}
\maketitle

\section{Introduction}

An open question in current AGN research is how the central black hole is
being fed with matter. The interstellar medium of the host galaxy seems to
be a good fuel source since gas and dust is found in large quantities
inside galaxies. What is not clear though, is how matter is transported
from galactic scales (kiloparsec scales) to the small scales (parsec
scales) near the nucleus.

One mechanism that has been proposed for the fueling of the active nucleus
is the existence of galactic bars (Shlosman 1990). Many statistical
studies have been performed in order to examine this possibility. Among
others, Xanthopoulos (1996) analyzed optical (V, R, I) observations
of a sample of 27 AGN and concluded that only half of them are barred.
Observations of AGN in the near-infrared (NIR) revealed that the fraction
of galaxies that host bars is larger than previously thought (e.g.
Mulchaey \etal 1997; M\'{a}rquez \etal 1999). This is mainly because dust
and star formation effects can mask the bar structures and thus make them
invisible at optical wavelengths while these effects are not very
important in the NIR wavelengths. Making use of NIR data, Mulchaey \&
Regan (1997) concluded that the incidence of bars in Seyfert and normal
galaxies is similar, suggesting that Seyfert nuclei do not occur
preferentially in barred galaxies.

Another mechanism that has been proposed for the fueling of AGN involves
interactions between the host galaxy and companion galaxies. It is argued
that a dynamical instability, caused by the tidal field of the companion
during a merger may drive a large fraction of the gas into the inner
regions of the galaxy (Hernquist 1989). Statistical studies have been made
to examine the environments of Seyfert and normal galaxies and the
percentage of companions around them. The results have been inconclusive
and rather ambiguous. For example, Fuentes--Williams \& Stocke (1988) and
Bushouse (1986) concluded that there is not a detectable difference in the
environments of Seyfert and normal galaxies. Dahari (1984) and Rafanelli
et al. (1995) on the other hand, found that Seyfert galaxies have an
excess of companions relative to normal galaxies.

Consequently, the observations so far have shown that bars and
interactions are not responsible for fueling the AGN in all cases.
However, irrespective of the mechanism that transports material close to
the central source, there should be enough potential fuel (in the form
of interstellar dust and gas) in the circumnuclear region of Seyfert
galaxies. Indeed, recent HST observations have shown that the central
regions in Seyfert galaxies are rich in gas and dust. Malkan \etal (1998)
found fine-scale structures of various morphological types in the central
regions of Seyfert 1 and 2 galaxies. Regan \& Mulchaey (1999) and Martini
\& Pogge (1999), examined the central morphology of Seyfert galaxies with
the use of color maps. They found significant structure, with gas and
dust organized mainly in nuclear spiral dust lanes on scales of a few
hundred parsec. These spiral dust lanes could be the channels by which the
material from the host galaxy is transported into the central engine.
Recently, Pogge \& Martini (2002) presented archival HST images of the
nuclear regions of Seyfert galaxies from the CfA Redshift Survey sample.
They found that essentially all the Seyfert galaxies in their sample have
circumnuclear structures which are connected with the large-scale bars and
spiral arms in the host galaxies and could be related to the fueling of
AGN by matter inflow from the host galaxy disks.

The studies mentioned above investigated the circumnuclear morphology in
AGNs mainly.  In the present work we compare the morphology of the central
regions in AGN and non-AGN galaxies in order to search for signatures of
the AGN fueling through differences that may exist. This comparison is
necessary in order to understand the main reason that makes a galaxy 
host an active nucleus. There are two obvious reasons that could cause
this effect, the first being the presence of a supermassive black hole in
the center of the galaxy, and the second is the presence of matter near
the central region which could provide the central black hole with the
necessary fuel. Since there is now sufficient observational evidence which
suggests that the majority (if not all) of the galaxies contain a central
black hole (e.g. Kormedy \& Gebhardt 2001), it is important to investigate
whether the lack of the fuelling material is the main reason for the
presence of nuclear activity.

We make use of HST observations of a group of nearby AGN and ``normal"
(i.e. non-AGN) galaxies which have similar distributions of distance,
morphological classification, as well as inclination. Using the ``ellipse
fitting" technique, we uncover their central, overall morphology. Any
localised excesses or deficits of emission (indicative of the presence of
significant amount of gas/dust) should show up as deviations from the
smooth isophotes. By measuring the average amplitude of these deviations
we can investigate, in a quantitative way, whether the central regions of
AGNs are significantly more ``irregular'' than those in normal galaxies.

The paper is organized as follows. In Sect. 2 we give information on the
sample of the galaxies, and in Sect. 3 we describe the method that we
follow in order to uncover the nuclear structure in the galaxies. In Sect.
4 we present our results. A discussion follows in Sect. 5, while a
summary of our work is presented in Sect. 6.

\section{The sample}

For the purposes of this study we chose objects from the Palomar optical
spectroscopic survey of nearby galaxies (Ho \etal 1995). These authors
surveyed a nearly complete sample of 486 bright ($B_{T}\leq 12.5$ mag),
northern ($\delta > 0^{0}$) galaxies using the Palomar 5m telescope and
derived a catalogue of emission-line nuclei, including a comprehensive
list of nearby AGNs (Ho \etal 1997a). The selection criteria of the survey
ensure that the sample is a fair representation of the local galaxy
population. Furthermore, the proximity of the objects enables fairly good
spatial resolution to be achieved, which is crucial for the objectives of
the present work.

Ho \etal (1997a) have classified the galaxies in their sample into various
subclasses of emission-line nuclei: H II nuclei, Seyfert nuclei, LINERs
and ``transition'' objects (i.e. composite LINER/H II nuclei). Although a
significant fraction of LINERs or transition objects could be genuine AGNs
(e.g. Ho 2001 and references therein) we considered only the classical
Seyfert 1s and 2s as representatives of the local AGN population. As
representatives of the non-AGN population we considered only the H II
galaxies and the galaxies that show no emission lines in their spectra. In
total, there are 52 Seyfert nuclei and 263 non-AGN galaxies in the Palomar
sample.

From this list of 315 objects we chose the galaxies that fulfiled the
following criteria: (1) inclination smaller than $70^{\circ}$, (2) they
were observed with the WFPC2 instrument onboard the HST (before the end of
1999) and their central region was mapped with the Planetary Camera (PC),
and (3) their central region was not overexposed. The small inclination
criterion was imposed because we are interested in studying the innermost
region of the galaxies which, in inclined systems, can be obscured due to
projection and/or obscuration effects. Most of the HST observations of the
Palomar sample have been performed with the WFPC2 instrument, hence, the
choice to study the galaxies which are observed with this instrument.
Finally, the requirement of PC observations of the central region was
enforced in order to maximize the available spatial resolution (the pixel
size of this camera is $0.0455\arcsec$, while the field of view is
$36\arcsec \times 36\arcsec$).

There are 58 galaxies (23 active and 35 non-active) that meet our
criteria.  Most of them were observed with the F555W and F606W filters (13
and 34, respectively), 10 objects were observed with the F547M filter, and
1 with the F569W filter. The difference between the effective wavelength
of the various filters is small and should not influence our results.
Table 1 lists the observational details (i.e. filter, exposure times and
ID program number of the original observing program) of the 58 galaxies.

\begin{table}
\begin{center}
\caption{Observational information of the galaxies}
\begin{tabular}{lccc} \hline
NAME & Filter & Exposure & Proposal ID \\
(NGC) &       & (sec)    &             \\ \hline
1058  &F606W  &80   & 5446 \\
1068  &F547M  &300  & 5479 \\
1358  &F606W  &500  & 5479 \\
1667  &F606W  &500  & 5479 \\
2273  &F606W  &500  & 5479 \\
2300  &F555W  &350  & 6099 \\
2639  &F606W  &500  & 5479 \\
2655  &F547M  &300  & 5419 \\
2748  &F606W  &400  & 6359 \\
2775  &F606W  &400  & 6359 \\
2903  &F555W  &400  & 5211 \\
2964  &F606W  &400  & 6359 \\
3031  &F547M  &100  & 5433 \\
3227  &F547M  &160  & 7403 \\
3310  &F606W  &500  & 5479 \\
3344  &F606W  & 80  & 5446 \\
3504  &F606W  &500  & 5479 \\
3516  &F547M  & 70  & 6416 \\
3810  &F606W  & 80  & 5446 \\
3982  &F606W  &500  & 5479 \\
4062  &F606W  & 80  & 5446 \\
4102  &F606W  &400  & 6359 \\
4138  &F547M  &200  & 6837 \\
4152  &F606W  &500  & 5479 \\
4168  &F547M  &230  & 6837 \\
4212  &F606W  & 80  & 5446 \\
4245  &F606W  & 80  & 5446 \\
4365  &F555W  &900  & 5920 \\
4371  &F606W  & 80  & 5446 \\
4378  &F606W  & 80  & 5446 \\
4379  &F555W  &160  & 5999 \\
4380  &F606W  & 80  & 5446 \\
4382  &F555W  &700  & 7468 \\
4405  &F606W  & 80  & 5446 \\
4406  &F555W  &500  & 5454 \\
4414  &F606W  & 80  & 8400 \\
4473  &F555W  &600  & 6099 \\
4477  &F606W  & 80  & 5446 \\
4478  &F555W  &400  & 6587 \\
4501  &F547M  &230  & 6837 \\
4536  &F555W  &300  & 5375 \\
4567  &F606W  & 80  & 5446 \\
4578  &F606W  & 80  & 5446 \\
4612  &F606W  & 80  & 5446 \\
4621  &F555W  &140  & 5512 \\
4639  &F547M  &230  & 5381 \\
4649  &F555W  &1100  & 6286 \\
4660  &F555W  &230  & 5512 \\
4694  &F606W  &500  & 5479 \\
4698  &F606W  &400  & 6359 \\
4800  &F606W  & 80  & 5446 \\
4900  &F606W  & 80  & 5446 \\
5033  &F547M  &230  & 5381 \\
5194  &F555W  &600  & 5777 \\
5273  &F606W  &400  & 8597 \\
6217  &F606W  &500  & 5479 \\
7479  &F569W  &600  & 6266 \\
7743  &F606W  &500  & 5479 \\
\hline
\end{tabular}
\end{center}
\end{table}

\begin{table}
\begin{center}
\caption{General and photometric properties of the galaxies}
\begin{tabular}{llccccc} \hline
NAME & Type & $T$ & D & $M_{B}$ & $i$ & Scale\\
(NGC) &       &   & (Mpc) &     & (deg) & (pc/pixel) \\ \hline
1058  &A(S2)  & 5  & 9.10  &-18.25  &21 &2.00 \\
1068  &A(S1.8) & 3  &14.40  &-21.32  &32 &3.17 \\
1358  &A(S2) & 0  &53.60  &-20.95  &38 & 11.80\\
1667  &A(S2) & 5  &61.20  &-21.52  &40 & 13.46\\
2273  &A(S2) & 0  &28.40  &-20.25  &41 & 6.25\\
2300  &NA  &-2  &31.00  &-20.69  &44 & 6.82\\
2639  &A(S1.9) & 1  &42.60  &-20.96 &54 & 9.37\\
2655  &A(S2) & 0  &24.40  &-21.12  &34 & 5.37\\
2748  &NA  & 4  &23.80  &-20.29  &70 & 5.25\\
2775  &NA  & 2  &17.00  &-20.34  &40 & 3.74\\
2903  &NA  & 4  & 6.30  &-19.89  &63 & 1.39\\
2964  &NA  & 4  &21.90  &-20.11  &58 & 4.82\\
3031  &A(S1.5) & 2  & 1.40  &-18.34 &60 & 0.31\\
3227  &A(S1.5) & 1  &20.60  &-20.39  &48 & 4.53\\
3310  &NA  & 4  &18.70  &-20.41  &40 & 4.11\\
3344  &NA  & 4  & 6.10  &-18.43  &24 & 1.34\\
3504  &NA  & 2  &26.50  &-20.61  &40 & 5.83\\
3516  &A(S1.2) &-2  &38.90  &-20.81  &40 & 8.56\\
3810  &NA  & 5  &16.90  &-20.19  &46 & 3.72\\
3982  &A(S1.9) & 3  &17.00  &-19.47 &30 & 3.74\\
4062  &NA  & 5  & 9.70  &-18.65  &67 & 2.13\\
4102  &NA  & 3  &17.00  &-19.54  &56 & 3.74\\
4138  &A(S1.9)&-1  &17.00  &-19.05  &50 & 3.74\\
4152  &NA  & 5  &34.50  &-20.34  &40 & 7.59\\
4168  &A(S1.9) &-5  &16.80  &-19.07  & .. & 3.70\\
4212  &NA  & 4  &16.80  &-19.78  &53 & 3.70\\
4245  &NA  & 0  & 9.70  &-17.92  &41 & 2.13\\
4365  &NA  &-5  &16.80  &-20.64  & .. & 3.70\\
4371  &NA  &-1  &16.80  &-19.51  &57 & 3.70\\
4378  &A(S2)& 1  &35.10  &-20.51  &21 & 7.72\\
4379  &NA  &-2  &16.80  &-18.60  &32 & 3.70\\
4380  &NA  & 3  &16.80  &-19.06  &58 & 3.70\\
4382  &NA  &-1  &16.80  &-21.14  &40 & 3.70\\
4405  &NA  & 0  &31.50  &-19.63  &51 & 6.93\\
4406  &NA  &-5  &16.80  &-21.39  & .. & 3.70\\
4414  &NA  & 5  & 9.70  &-19.31  &57 & 2.13\\
4473  &NA  &-5  &16.80  &-20.10  & .. & 3.70\\
4477  &A(S2) &-2  &16.80  &-19.83  &24 & 3.70\\
4478  &NA  &-5  &16.80  &-18.92  & .. & 3.70\\
4501  &A(S2) & 3  &16.80  &-21.27  &59 & 3.70\\
4536  &NA  & 4  &13.30  &-20.04  &67 & 2.93\\
4567  &NA  & 4  &16.80  &-19.34  &48 & 3.70\\
4578  &NA  &-2  &16.80  &-18.96  &43 & 3.70\\
4612  &NA  &-2  &16.80  &-18.75  &38 & 3.70\\
4621  &NA  &-5  &16.80  &-20.60  & .. & 3.70\\
4639  &A(S1) & 4  &16.80  &-19.28  &48 & 3.70\\
4649  &NA  &-5  &16.80  &-21.43  & .. & 3.70\\
4660  &NA  &-5  &16.80  &-19.06  & .. & 3.70\\
4694  &NA  &-2  &16.80  &-19.08  &63 & 3.70\\
4698  &A(S2) & 2  &16.80  &-19.89  &53 & 3.70\\
4800  &NA  & 3  &15.20  &-18.78  &43 & 3.34\\
4900  &NA  & 5  &17.30  &-19.10  &21 & 3.81\\
5033  &A(S1.5) & 5  &18.70  &-21.15  &64 & 4.11\\
5194  &A(S2) & 4  & 7.70  &-20.76  &53 & 1.69\\
5273  &A(S1.5) &-2  &21.30  &-19.26  &24 & 4.69\\
6217  &NA  & 4  &23.90  &-20.23  &34 & 5.26\\
7479  &A(S1.9)& 5  &32.40  &-21.33  &41 & 7.13\\
7743  &A(S2) &-1  &24.40  &-19.78  &32 & 5.37\\
\hline
\end{tabular}
\end{center}
\end{table}

In Table 2 (columns 3, 4, 5 and 6) we list the global and photometric
properties of the 58 galaxies, i.e. the numerical Hubble type index ($T$),
their distance, the absolute $B$ band magnitude ($M_{B}$), corrected for
intrinsic and galactic absorption, and the inclination ($i$) (the data
listed in this table were taken from Ho \etal, 1997a). The parameter $T$
ranges from -5 to 5 with 25 galaxies being early-type ($T\le0$) and 33
galaxies being late-type ($T>0$). Column 2 in Table 2 lists the galaxy
``activity type'': ``A'' and ``NA'' stand for AGN and non-AGN, according to
the classification of Ho \etal (1997a) while for AGNs the
Seyfert type is also given. Finally, the last column in the
same table lists the projected scale in parsecs per PC pixel at the
distance of the galaxy.

Although the original sample of Ho \etal (1995) is an almost complete
sample of nearby galaxies (see discussion in Ho \etal, 1997b), this is not
true for the present sample. Since the main selection criterion is the
availability of WFPC2 observations, it is important to examine if sample
biases and selection effects are introduced in this way.

\begin{figure}
\psfig{figure=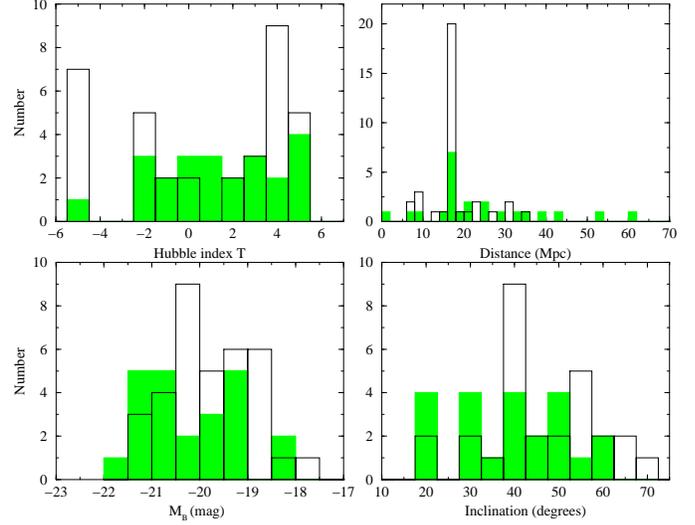,width=9.0truecm,angle=-90,%
 bbllx=0pt,bblly=50pt,bburx=580pt,bbury=720pt}
\caption[]{Distribution of the Hubble type index, distance, absolute B band
magnitude, and inclination for the AGN and non-AGN group of galaxies
(filled and open histograms, respectively).}
\label{f1}
\end{figure}

Table 3 lists the median global properties of the objects in our sample
and Fig.~\ref{f1} shows the distribution of morphological type, distance,
absolute magnitude and inclination for the AGN and non-AGN group of
galaxies (filled and open histograms, respectively). The average
properties of the galaxies are similar for both groups (Table 3). The
exception is the absolute magnitude, with the AGNs being
brighter than the non-AGNs by $\sim 0.7$ mags on average.

Application of the Kolmogorov--Smirnov (K--S) test (Press \etal 1992)  
confirms that the distributions plotted in Fig.~\ref{f1} are similar,
apart from the distributions of $M_{B}$. The probability that the sample
distributions of $T$, distance and inclination are drawn from the same
parent population is $78\%$, $15\%$, and $37\%$, respectively (hereafter,
when we compare different distributions or compute correlation
coefficients, we consider as ``statistically significant'' these
differences or correlations with a probability to appear by chance being
less than $10\%$). In the case of the distributions of $M_{B}$, the K--S
test gives a probability of only $5\%$. We conclude that the distributions
of distance, morphological classification and inclination of the non-AGN
sample match those of the AGN sample. The distribution of the galaxy
luminosity is different between the two samples, however, as we discuss in
Sect. 4 this does not affect our conclusions.

\begin{table}
\begin{center}
\caption{Median Properties of the active and non-active group of galaxies}
\begin{tabular}{lccccc} \hline
Group & $T$ & D & $M_{B}$ & $i$ & Scale \\ 
      &   & (Mpc) &  & (deg)  & (pc/pixel) \\ \hline
A & 1 & 17.8 & $-20.5$ & 40 & 3.74\\
NA & 2 & 16.8 & $-19.8$ & 44 & 3.70\\
\hline
\end{tabular}
\end{center}
\end{table}  

\section{The Data and method of analysis}

All the WFPC2 images that we used were already processed by the standard
STScI reduction pipeline (Biretta \etal 2000). Since we have chosen images
with no saturation effects, the only additional processing step required
is the removal of cosmic rays. For that reason, we used the {\it
filter/cosmic} task of MIDAS astronomical package. Any residual cosmic ray
events as well as bright foreground stars were removed by hand. For those
galaxies with more than one image, we chose to study the one with the
largest exposure time which did not cause saturation effects in the
central region. If there were two or more images of the same integration
time, we analyzed the images (as explained below) and then combined the
resulting ``variance'' values (see below).

Our main aim is to study, in a quantitative way, the irregularities in the
morphology of the central region of the galaxies. Our methodology consists
of two steps. In the first step, we use the ellipse fitting technique to
recover the axisymmetric isophotes around the center of the galaxies. If
there are localized regions with excess emission (for example regions of
star clusters or H II regions) or deficits (caused by dust absorption),
they will show as deviations from the smooth isophotes. Based on this
idea, in the second step, we compute the scatter of the pixel values
around their mean (i.e. their variance) and use this value as a measure of
the amplitude of the central structures. Although it is hard to estimate
the significance of the derived amplitudes for each individual galaxy, the
comparison of the distribution of the amplitudes for various groups of
galaxies can provide us with useful information. For example, one would
expect that AGNs, hosting a black hole and a large amount of gas in their
nuclear region should show larger amplitude structure (and thus larger
variance), on average, when compared with the normal galaxies. We describe
below the two steps in more detail.

\begin{figure*}
\psfig{figure=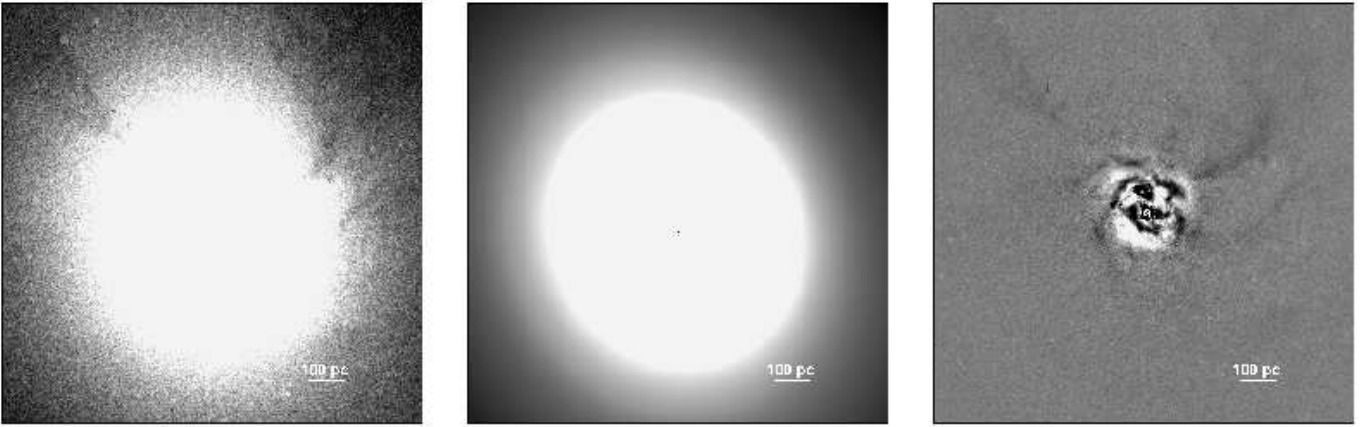,width=14.5truecm,angle=0,%
 bbllx=-90pt,bblly=540pt,bburx=545pt,bbury=270pt}
\caption[]{HST image of the galaxy NGC 5273 (left panel). The
artificial image from the fitted ellipses is shown on the middle panel,
while the image that is produced after dividing the HST image with
the artificial image is shown in the right panel. The white bar
in all these images shows the distance scale.}
\label{f2}
\end{figure*}

\begin{figure*}
\psfig{figure=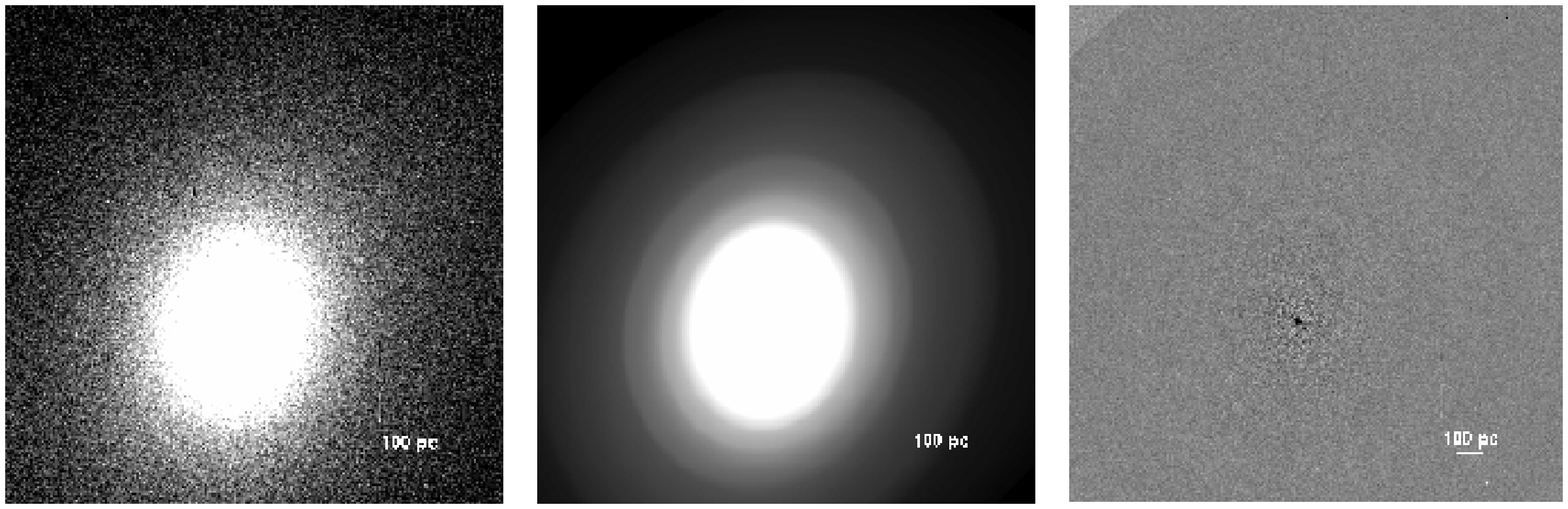,width=15.2truecm,angle=0,%
 bbllx=-40pt,bblly=540pt,bburx=545pt,bbury=270pt}
\caption[]{HST image of the galaxy NGC 4612 (left panel). The
artificial image from the fitted ellipses is shown on the middle panel,
while the image that is produced after dividing the HST image with
the artificial image is shown in the right panel. The white bar
in all these images shows the distance scale.}
\label{f3}
\end{figure*}

\begin{figure*}
\psfig{figure=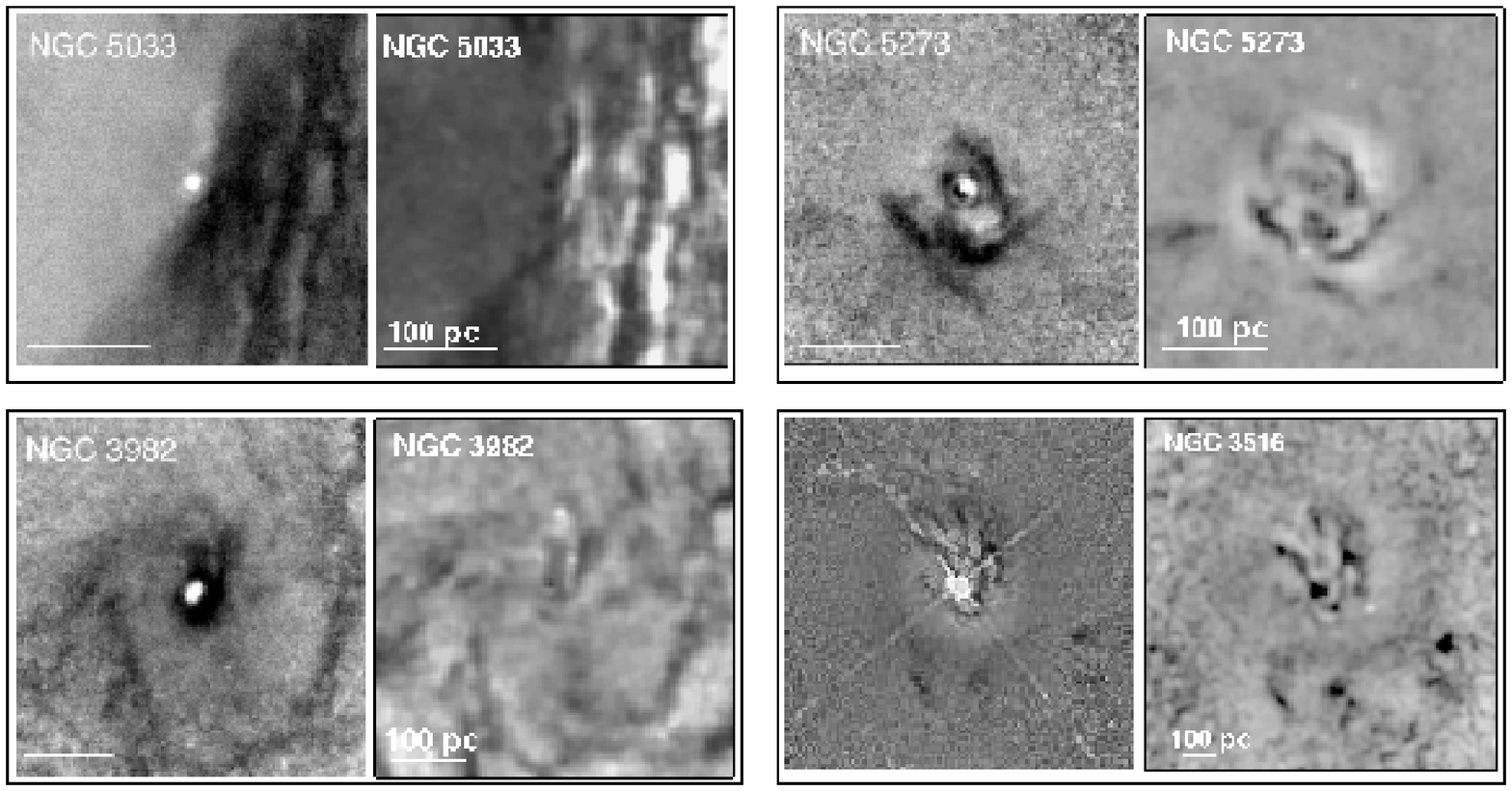,width=17.9truecm,angle=0,%
 bbllx=15pt,bblly=550pt,bburx=600pt,bbury=250pt}
\caption[]{``Structure'' maps of the galaxies NGC 5033,
NGC 5273, NGC 3982 and NGC 3516 (top left, top right, bottom
left and bottom right respectively). The maps that were created
with the ellipse fitting technique (right hand side on each
panel) are compared with the maps taken from the literature
(left hand side on each panel; courtesy of Martini \& Pogge
(1999) for NGC 5033, NGC 5273 and NGC 3982 and courtesy of
Pogge \& Martini (2002) for NGC 3516). The white bar in the
images indicates the distance scale.}

\label{maps}
\end{figure*}

As mentioned above, first, we perform ellipse fitting to the isophotes of
the galaxy image.  This choice is motivated by the fact that the isophotes
of galaxies, especially elliptical (E) and lenticular (S0) as well as the
bulge of spiral galaxies, are not far from ellipses. This technique has
been widely used in the past by various authors, mainly as a method of
retrieving embedded galaxy structures that are hidden by the large-scale
distribution of light of the main body of the galaxy. Descriptions of
ellipse fitting techniques and their applications to the surface
photometry of galaxies can be found in Kent (1983), Jedrzejewski (1987),
Bender \& M\"{o}llenhoff (1987), Wozniak et al. (1995) and Milvang--Jensen
\& J\o rgensen (1999). Using the {\it fit/ell3} task of MIDAS, which is
based on the formulas of Bender \& M\"{o}llenhoff (1987), we fit the
isophotes of the galaxy image and construct an artificial image from the
fitted ellipses.

As an example, in Fig.~\ref{f2} we show the central region of NGC 5273
(left panel) together with the ``artificial'' image from the fitted
ellipses (middle panel).  In order to recover the morphological features
in the innermost part of the galaxy we then divide the galaxy image by the
artificial image from the fitted ellipses. In doing so, we normalise the
pixel values of the central regions in all galaxies to unity so that the
amplitude of the structures in one galaxy can be compared with the
amplitude in other galaxies, irrespective of their brightness or exposure
time. The resulting image for NGC 5273, which we call the ``structure''
image, is also shown in Fig.~\ref{f2} (right panel). If there were no
structure in the central region, we would expect a smooth image with all
the pixels having a value around $\sim 1$. On the contrary, the image on
the right panel of Fig.~\ref{f2} shows positive and negative deviations
from the smooth isophotes, indicative of localised structures. As an
example of a galaxy with no deviations from the underlying galactic
isophotes, in Fig.~\ref{f3} we show the HST image of NGC 4612, together
with the image from the fitted ellipses and the resulting image, after
dividing the HST image with the image from the elliptical isophotes (left,
middle and right panel, respectively). The structure image is almost
completely smooth, with an average value of $1.00 \pm 0.05$ and no obvious
deviations from the fitted isophotes.

In the second step, we choose two regions around the center of each galaxy
with size of 100 pc and 1 kpc. The regions are defined as orthogonal boxes
centered on the galaxy and with their largest side parallel to the major
axis of the galaxy. The ratio of the small-to-large side of the box is
taken as the cosine of the inclination angle of the galaxy. First, we
measure the mean of the pixel values in each region and then their
variance by using the {\it statistics/image} task of MIDAS. There are two
mechanisms that contribute to the variance that we measure. One is the
photon noise process and the other is the presence of 
features/irregularities in the region. Since we are interested in
measuring the amplitude of the deviations that are caused by the galaxy
micro-structures, we have to correct the estimated variance for the
contribution of the photon noise statistics. Let us denote with
$\sigma^{2}_{R}$, $\sigma^{2}_{PN}$, and $\sigma^{2}_{S}$ the total
variance of a region, say $R$, the variance that is introduced by the
photon noise process, and the variance that is due to micro-structures,
respectively. Since the photon noise variations and the variations due to
morphological irregularities contribute to $\sigma^{2}_{S}$ in an
independent way, then $\sigma^{2}_{R}=\sigma^{2}_{PN}+\sigma^{2}_{S}$, or
$\sigma^{2}_{S}=\sigma^{2}_{R}-\sigma^{2}_{PN}$. Therefore, we have to
subtract the variance that is caused by the photon noise statistics from
the total variance of each region in order to compute the variance that is
due to real galactic structures. This is an important correction, since
the flux (and hence the photon noise statistics) is different for each
galaxy due to differences in the exposure time and their magnitude.

The contribution of each pixel to the total variance in each region is
$(x_{i}-\bar{x})^{2}/N$, where $x_{i}$ is the pixel value, $\bar{x}$ is
the mean value of all the pixels in the region and $N$ is the total number
of pixels. The photon noise process contribution for that pixel is
$(x_{i}g+RON^{2})/N$, where $g$ and $RON$ are the gain and read-out noise
of WFPC2 ($7e^-/DN$, and $5e^-$), respectively. Therefore, the
contribution to the total variance of the photon noise statistics is
$\sigma^{2}_{PN}=\sum_{i}(x_{i}g+RON^{2})/N$, where the summation is over
the $N$ the pixels of the region. In order to estimate this, we
constructed an ``error image'', i.e. we multiplied the original image with
$g$, added the $RON^2$ value, and then divided with $(m_{i}g)^2$ in order
to take account for the fact that we have also divided the original image
with the fitted ellipses ($m_{i}$ is the value of the elliptical isophot
at each pixel $x_{i}$). The mean value of the regions in the ``error"
image which have the same dimensions as the respective regions in the
structure image is a good approximation of $\sigma^{2}_{PN}$.

\section{Results}

The estimated values of $\sigma^{2}_{S}$ for the inner 100 pc and 1 kpc
regions of all the galaxies in our sample are listed in Table 4 (columns 2
and 3, respectively). In effect, the variance that we estimate gives a
measure of the average amplitude of the deviations from the smooth
isophotes in each galaxy. The average $\sigma^{2}_{S}$ in the innermost
100 pc and 1 kpc regions is $0.06\pm 0.02$ and $0.025\pm 0.006$,
respectively (note that for the nearest galaxies, the 1 kpc region is
larger than the field of view of the PC camera; for that reason we could
not estimate their $\sigma^{2}_{S1kpc}$). Consequently, the average
amplitude of the localised structures in the innermost 100 pc and 1 kpc
regions are $\sqrt{\sigma^{2}_{S}}\times100\% \sim 25\%$ and $\sim 16\%$
of the underlying galaxy emission, respectively. The fact that the average
structure amplitude {\it decreases} with distance from the center (i.e.
$\sigma^{2}_{S1kpc} < \sigma^{2}_{S100pc}$) implies that most of the
structure is concentrated at the center of the galaxies. As a result,
consideration of a larger region will tend to dilute the signal, i.e.
decrease $\sigma^{2}_{S}$.

Before comparing the structure amplitudes for the various groups of
galaxies, we have to examine whether our results are biased by any
observational or global characteristics of the galaxies. First of all, if
the central region isophotes are not elliptical then the residuals that we
detect could be the result of the failure of the elliptical isophotes to
fit properly the underlying starlight distribution. An indication that the
``ellipse fitting" method works successfully in suppressing the underlying
galaxy distribution and revealing real structures in the central regions
of the galaxies is given by the fact that most of the dust/emission
structures that we detect are certainly visible in the original images as
well. In order to investigate further this possibility, we plotted the
fitted ellipses superimposed on the galactic isophotes for all the
galaxies in our sample. For galaxies with $T\le0$, the agreement between
the overall shape of the isophotes and the fitted ellipses is very good.
In many cases the isophotes are not smooth and small-scale departures from
the elliptical shape are apparent, indicative of the localised structures
that we want to study. For the $T>0$ galaxies, in some cases, we do
observe systematic deviations of the isophotes from ellipses at large
distances from the galactic centers. They are caused mainly by the
presence of spiral arms, while in a few cases the presence of an inclined
disk causes the isophotes to become more elongated than the fitted
ellipses. However, at small radii, the isophotes are well approximated by
ellipses, with any deviations being localised and suggestive of
small-scale structures.  

As a final test of how well the ellipse fitting method detects the
underlying morphological signatures, we compared our ``structure'' images
with images obtained by using other techniques. In Fig.~\ref{maps}, we
plot the ``structure" image of the Seyfert 2 galaxies NGC 5033, NGC 5273
and NGC 3982 (top left, top right and bottom left respectively; note that
the structure image of NGC 5273, also presented in Fig. 2, is now fliped
and scaled to different brightness levels so that it can be compared with
the respective color map image). The central micro-structure in these
galaxies have been studied by Martini \& Pogge (1999) with the use of
($V-H$) color maps. These maps are shown in Fig.~\ref{maps} as well. In
the same figure, we also plot the ``structure" image of NGC 3516 (a
Seyfert 1 galaxy; bottom right panel) together with another structure
image which was constructed with the use of the ``Richardson - Lucy''
(R--L) image restoration technique (Pogge \& Martini, 2002). In all
panels, the left hand side images are those taken from the literature
(downloaded in electronic form from the respective journal site) and the
right hand side images are the ``structure'' images created with the
ellipse fitting method presented in this work. The brightness levels and
the image scales were adjusted in such a way so that they are roughly
comparable with the ones taken from the literature.

Comparison of the ``structure" images with the color maps in
Fig.~\ref{maps} shows clearly that the ellipse fitting technique reveals
successfully most of the features that appear in the color maps (in some
cases, the features appear more enhanced in the ``structure" images).
There are no additional features that could be identified with ellipse
fitting residuals. It is also evident that the bright cores seen in all
the galaxies in the color maps are fitted quite well with ellipses. As a
result, they do not appear and do not cause any artifacts in the
``structure'' images. This is true even in the case of NGC 3516. In order
to avoid the overexposed bright nucleus which appears in the image of
Pogge \& Martini (2002), we have used an image of shorter exposure, taken
with a different filter. The nucleus is successfully removed, and the same
features appear in both images. We conclude that the ``ellipse fitting''
method works successfully in suppressing the underlying smooth galaxy
distribution and revealing real structures in the central regions of the
galaxies. However, it is hard to judge whether the $\sigma^{2}_{S}$ values
(of the $T>0$ galaxies mainly) are indicative of the real structure
amplitudes {\it only}, or whether the amplitude of any fitting method
residuals contributes significantly as well. Because of this reason, the
$\sigma^{2}_{S}$ values should be considered as a rough estimate of the
galactic micro-structure amplitudes.

\begin{figure}
\psfig{figure=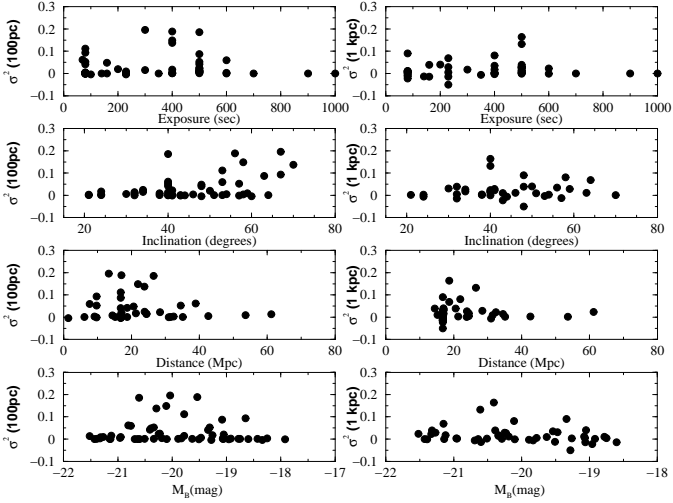,width=8.5truecm,angle=-90,%
 bbllx=0pt,bblly=50pt,bburx=580pt,bbury=720pt}
\caption[]{Plot of the estimated variance for the 100 pc and 1 kpc inner
region of the galaxies as a function of exposure time, inclination,
distance and absolute $B$ band magnitude (from top to bottom,
respectively). Note that for clarity reasons, the two galaxies with the
highest $\sigma^{2}_{S 100pc}$ values are not plotted, although their
values are considered in the estimation of the correlation coefficients
as discussed in the text.}
\label{f4}
\end{figure}

Furthermore, Fig.~\ref{f4} (upper panel) shows a plot of $\sigma^{2}_{S}$
as a function of exposure time, for both the 100 pc and 1 kpc regions.
Although we have normalised the structure images to the underlying galaxy
isophotes, hence the estimation of $\sigma^{2}_{S}$ is not affected by
differences in the brightness of the galaxies, there is still the
possibility that if the signal to noise is small in some cases (due to
short exposure time) we may not be able to estimate accurately the
amplitude of the central structures. As Fig.~\ref{f4} shows, this is not
the case. There is no correlation between $\sigma^{2}_{S100 pc}$ or
$\sigma^{2}_{S1kpc}$ with exposure time. This result is verified when we
use Kendall's $\tau$ nonparametric statistic (Press \etal 1992) in order
to investigate, quantitatively, whether there is a significant correlation
between the two variables. We find $\tau=0.05$ and $\tau=-0.15$ for the
[$\sigma^{2}_{S100pc}$, exposure] and [$\sigma^{2}_{S1kpc}$, exposure]
variables, respectively. The probability that we would obtained these
values by chance, if the two variables were uncorrelated, is $\sim 15\%$
in both cases.

On the other hand, both $\sigma^{2}_{S100pc}$ and $\sigma^{2}_{S1kpc}$ are
correlated with the inclination and distance of the galaxies. Looking at
Fig.~\ref{f4} (second panel from top) we can see a positive correlation
between $\sigma^{2}_{S}$ and inclination: as the inclination increases, so
does $\sigma^{2}_{S}$. The $\sigma^{2}_{S}$ vs distance plot (second panel
from bottom in Fig.~\ref{f4}) shows that $\sigma^{2}_{S}$ is also
correlated with distance. In fact, in this case, an anti--correlation is
observed. As the distance decreases, $\sigma^{2}_{S}$ increases.
Computation of Kendal's $\tau$ yielded 0.28, 0.24, -0.16 and -0.23 for the
$\sigma^{2}_{S100pc, S1kpc}$ vs inclination, and the $\sigma^{2}_{S100pc,
S1kpc}$ vs distance correlations, respectively. The probability that these
values would appear by chance if the variables [$\sigma^{2}_{S}$,
inclination] and [$\sigma^{2}_{S}$, distance] were uncorrelated is $0.2\%$
and $8\%$ (in the case of the $\sigma^{2}_{S100pc}$ plots) and $2\%$,
$3\%$ (in the case of the $\sigma^{2}_{S1kpc}$ plots) respectively. The
dependence of $\sigma^{2}_{S}$ on distance is easy to interpret. The
median distance of all the galaxies is 16.7 Mpc. However, the distance of
the nearest galaxy is only 1.4 Mpc, while the most distant galaxy is
located at 61.2 Mpc. Any small scale structures will be smoothed out in
the more distant galaxies, hence the increase of $\sigma^{2}_{S}$ with
decreasing distance. The dependence of the variance on the inclination is
rather unexpected. One would expect that small scale structure would be
diminished in inclined systems, while the opposite effect is observed.
Visual inspection of the respective structure images shows that the
increase of $\sigma^{2}_{S}$ with increasing inclination is caused by the
obscuration effects due to dust, which become more pronounced in inclined
systems.

Finally, we also examined whether $\sigma^{2}_{S}$ depends on the absolute
$B$ band magnitude of the galaxies (Fig.~\ref{f4}, lower panel). As
expected, there seems to be no correlation between the two variables.
Indeed, $\tau=-0.07$ and $-0.15$ for the [$\sigma^{2}_{S100 pc, 1kpc},
M_{B}$] variables. The probability that the correlation of two
uncorrelated variables would yield these values by chance is $41\%$ and
$15\%$ respectively.

\begin{table}
\begin{center}
\caption{The microstructure variance of the central regions of the
galaxies}
\begin{tabular}{lcc} \hline
NAME & $\sigma^{2}_{S}$ & $\sigma^{2}_{S}$\\
(NGC) &   100 pc    &  1 kpc    \\ \hline
1058  &  0.00300  &  ..\\
1068  &  0.00810  &  0.03900\\
1358  &  0.00920  &  0.00200\\
1667  &  0.01300  &  0.02400\\
2273  &  0.02260  &  0.02900\\
2300  &  0.00008  & -0.00570\\
2639  &  0.00510  &  0.00320\\
2655  &  0.01500  &  0.01700\\
2748  &  0.13700  &  0.00060\\
2775  &  0.00070  &  0.01700\\
2903  &  1.16400  &  ..\\
2964  &  0.14900  &  0.08100\\
3031  & -0.00440  &  ..\\
3227  &  0.04800  &  0.03900\\
3310  &  0.04200  &  0.16400\\
3344  &  0.00130  &  ..\\
3504  &  0.18600  &  0.13200\\
3516  &  0.06200  &  ..\\
3810  &  0.00420  &  0.01100\\
3982  &  0.00600  &  0.03100\\
4062  &  0.09340  &  ..\\
4102  &  0.18900  &  0.03500\\
4138  &  0.02000  &  0.04000\\
4152  &  0.05200  &  0.01400\\
4168  &  0.00250  &  0.00510\\
4212  &  0.11200  &  ..\\
4245  & -0.00120  &  ..\\
4365  & -0.00008  & -0.00050\\
4371  & -0.00120  & -0.01200\\
4378  &  0.00080  &  0.00160\\
4379  & -0.00020  & -0.01460\\
4380  &  0.00450  &  ..\\
4382  &  0.00026  & -0.00036\\
4405  &  0.00130  &  0.00900\\
4406  &  0.00090  & -0.00090\\
4414  &  0.05200  &  ..\\
4473  & -0.00006  & -0.00060\\
4477  &  0.00310  & -0.00630\\
4478  &  0.00180  &  0.00070\\
4501  &  0.00930  &  0.02900\\
4536  &  0.19600  &  ..\\
4567  &  0.04100  &  0.09000\\
4578  & -0.00070  & -0.02280\\
4612  &  0.00100  &  0.00400\\
4621  &  0.00017  & -0.01370\\
4639  & -0.00400  & -0.05000\\
4649  & -0.00006  & -0.00020\\
4660  & -0.00040  & -0.01400\\
4694  &  0.08700  &  0.01100\\
4698  &  0.00080  & -0.00300\\
4800  &  0.00090  &  0.01000\\
4900  &  0.91000  &  ..\\
5033  &  0.00140  &  0.06900\\
5194  &  0.05900  &  ..\\
5273  &  0.01780  &  0.00300\\
6217  &  0.02500  &  0.02700\\
7479  &  0.00260  &  0.02300\\
7743  &  0.01420  &  0.00492\\
\hline
\end{tabular}
\end{center}
\end{table}

\subsection{Active vs non-Active galaxies}

\begin{figure}
\psfig{figure=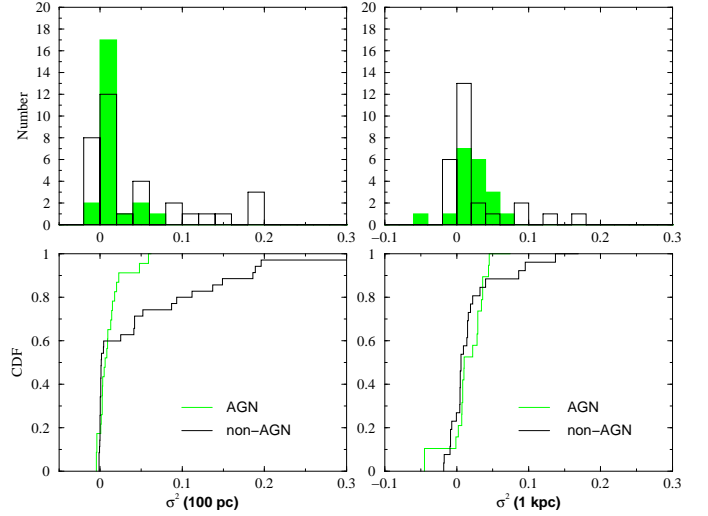,width=9.0truecm,angle=-90,%
 bbllx=0pt,bblly=50pt,bburx=580pt,bbury=720pt}
\caption[]{ Distributions and cumulative distribution functions of
$\sigma^{2}_{S100pc}$ (left panels), and $\sigma^{2}_{S1kpc}$ (right
panels) for the active and non-active group of galaxies (filled and open
histograms, respectively, for the distributions). Note that the two
non-active galaxies with the highest $\sigma^{2}_{S100pc}$ values are not
plotted in the left, upper panel, for clarity reasons.}
\label{f5}
\end{figure}

In Fig.~\ref{f5} we plot the distribution of $\sigma^{2}_{S100pc}$ and
$\sigma^{2}_{S1kpc}$ for the AGN and non-AGN galaxies in our sample
(filled and open histograms, respectively). The two distributions appear
to be similar, although the non-AGN galaxies show an excess of larger
$\sigma^{2}_{S100pc}$ values. This effect is more pronounced in the plots
of the cumulative distribution functions (CDF) of $\sigma^{2}_{S}$ (shown
in the lower panel of Fig.~\ref{f5}). The $\sigma^{2}_{S100pc}$ CDF of the
non-AGN galaxies shows clearly an extended tail towards larger values. The
K--S test shows that the differences between the two distributions are not
statistically significant. The probability that they are drawn from the
same population is $11\%$ and $12\%$ for the $\sigma^{2}_{S100pc}$ and
$\sigma^{2}_{S1kpc}$ distributions respectively. We note that the
dependence of $\sigma^{2}_{S}$ on distance and inclination cannot
influence our results since the distributions of inclination and distance
are statistically similar for the active and non-active groups of
galaxies.

\begin{figure}
\psfig{figure=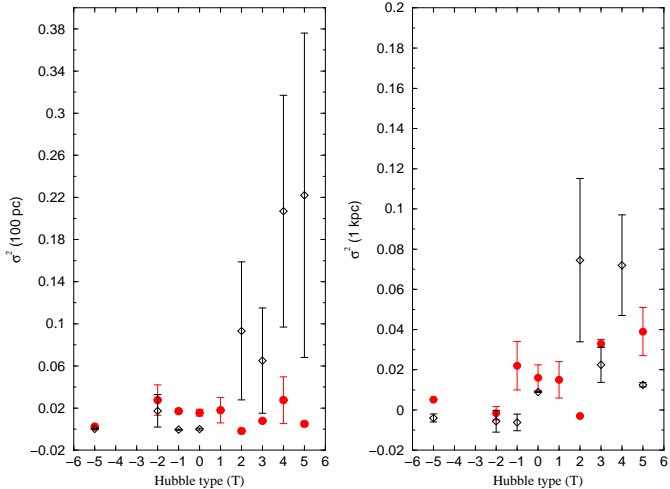,width=9.0truecm,angle=-90,%
 bbllx=0pt,bblly=20pt,bburx=580pt,bbury=720pt}
\caption[]{Plot of the average variance of the AGN and non-AGN groups
of galaxies as a function of $T$ (filled circles and open diamonds,
respectively). Each point represents the average variance of all
the galaxies with the same $T$ value. The errorbars show the error on the
mean, and are absent in the case when there is only one galaxy in a
$T$ bin.}
\label{f6}
\end{figure}

In order to examine whether $\sigma^{2}$ depends on the galaxy Hubble type
and the galaxy activity class, in Fig.~\ref{f6} we plot the mean
$\sigma^{2}_{S100pc}$ and $\sigma^{2}_{S1kpc}$ for the AGN and normal
galaxies as a function of $T$. In the $\sigma^{2}_{S1kpc}$ vs $T$ plot we
observe a systematic increase of the structure amplitude with the Hubble
type, which is roughly similar for both the AGNs (filled circles) and
non-AGNs (open diamonds). This trend is expected, for three reasons. At a
distance of 1 kpc, as $T$ increases, we start to detect the increasing
amplitude of the spiral arms (with respect to the underlying
galactic/bulge component). Furthermore, early-type galaxies are less
gas--rich than late-type galaxies (Young \& Scoville 1991). Therefore,
obscuration effects and/or bright H II regions will become more prominent
(and hence the variance will increase) as $T$ increases. Finally, as we
mentioned in Section 4, in late-type galaxies, the presence of an
exponential disk can cause fitting residuals to appear, and hence increase
the $\sigma^{2}_{S1kpc}$ values for these galaxies. 

The $\sigma^{2}_{S100pc}$ vs $T$ plot shows a similar trend (the variance
increases with increasing $T$), however there are two important
differences as well. Firstly, AGNs appear to have similar
$\sigma^{2}_{S100pc}$ values, irrespective of their Hubble type. On the
other hand, the early and late normal galaxies show a large difference in
their $\sigma^{2}_{S100pc}$, with the late-type galaxies being much more
irregular in their central regions than the early-type galaxies. We would
like to emphasize that the differences in the variance of the early and
late-type galaxies can not be caused by differences in their distance or
inclination, since the distributions of the distance and inclination for
the two groups of galaxies are statistically similar (the probability that
they are drawn from the same parent population is $27\%$ and $12\%$,
respectively), or by any fitting method residuals, since the overall shape
of the galactic isophotes at small radii in both early and late-type
galaxies are well approximated by ellipses as we mentioned in Section 4.

\subsection{Active vs non-Active early and late-type galaxies}

\begin{figure}
\psfig{figure=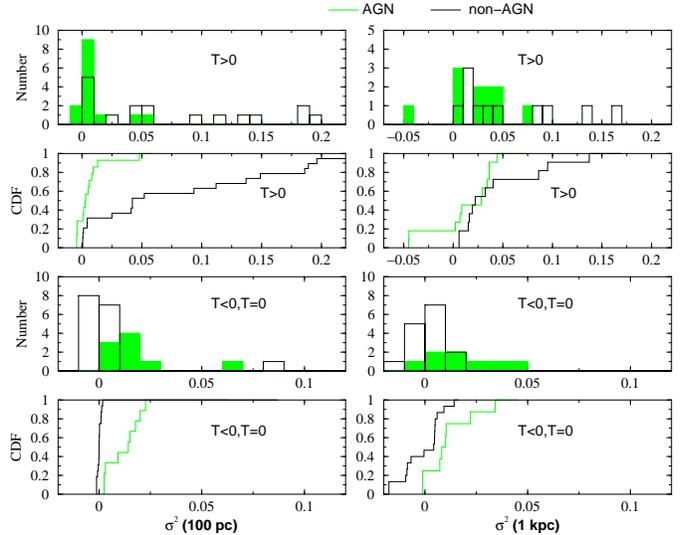,width=9.0truecm,angle=-90,%
 bbllx=0pt,bblly=50pt,bburx=580pt,bbury=720pt}
\caption[]{Distributions and CDFs of the variance [both for the inner 
100 pc (left) and 1 kpc (right) regions] for AGN (filled histograms) and
non-AGN (open histograms) galaxies, which are late-type (upper two panels)
and early-type (lower panels). In the upper left panel, the two non-AGN
galaxies with large variance values are not plotted for clarity reasons.}
\label{f7}
\end{figure}

We investigated further the comparison between the $\sigma^{2}$
distributions of the AGN and non-AGN groups of galaxies taking into
account the differencies between the early and late-type of galaxies that
Fig.~\ref{f6} revealed. In Fig.~\ref{f7}, we plot the distributions of
$\sigma^{2}$ for AGN and non-AGN considering not the whole samples but
only late or early-type galaxies (upper and second from bottom panel,
respectively). In the case of late-type galaxies, the distribution of
$\sigma^{2}_{S}$ for normal galaxies extends to much larger values, when
compared to the distribution of the AGN. The opposite is true in the case
of early-type galaxies. There, the AGN distribution of $\sigma^{2}_{S}$
shows a tail towards larger values, while the distribution of the normal
galaxies is centered around $\sim 0$.

These results become more pronounced in the plots of the cumulative
distribution functions. The CDF plot of $\sigma^{2}_{S100pc}$ for
late-type AGN and non-AGN shows clearly that, on average, normal galaxies
have $\sigma^{2}_{S}$ values larger than the AGN (second from top panel in
Fig.~\ref{f7}). On the other hand, the CDF of the early-type AGN is
shifted to larger values when compared to the CDF of the early type
non-AGN galaxies (bottom panel in Fig.~\ref{f7}). Application of the K--S
test shows that the distributions shown in Fig.~\ref{f7} are significantly
different. The probability that they are drawn from the same parent
population is less than $0.1\%$ in the case of early-type galaxies (for
both $\sigma^{2}_{S100pc}$ and $\sigma^{2}_{S1kpc}$), and $0.5\%$ in the
case of the distributions of $\sigma^{2}_{S100pc}$ for late-type AGN and
non-AGN. On the other hand, the $\sigma^{2}_{S1kpc}$ distributions for the
same group of galaxies, are statistically similar, with a probability of
being drawn from the same parent population of $37\%$.

The difference between the distributions of $\sigma^{2}_{S}$ of the
non-AGN/AGN early and late type galaxies cannot be caused from differences
in the distributions of their morphological type or inclination. For
example, application of the K--S test shows that the Hubble type index and
the inclination distributions of the early type, AGN and non-AGN galaxies
are {\it not} significantly different. The probability of being drawn from
the same parent polution is $48\%$ and $31\%$ respectively.  The only
significant difference that we find is between the distributions of
distance for the early-type AGN/non-AGN galaxies. AGN have on average
larger distance compared to the non-AGN, early-type galaxies. However,
this result implies that the difference in the variance between the two
groups is actually {\it larger} than what is observed in Fig.~\ref{f7}.

\section{Discussion}

We have used archival WFPC2 HST images to study the morphology of 58 nearby
galaxies in a quantitative way, i.e. by measuring the variance (that is to
say the average amplitude of the deviations from the smooth galactic
isophotes), in two regions around the center of each galaxy, one with a
radius of 100 pc and the other with a radius of 1 kpc. Our main results are
as follows:

1) Taken as a whole, the galaxies show considerable structure in their
central regions. The amplitude of the nuclear features is roughly $\sim
25\%$ and $\sim 15\%$ of the underlying galactic emission in the inner
region of 100 pc and 1 kpc, respectively.

2) When we consider the whole AGN and non-AGN group of galaxies, the
distributions of the variances are statistically similar.

3) The central structure tends to increase from early-type towards
late-type galaxies. 

4) The $\sigma^{2}_{S1kpc}$ increases ``smoothly'' for both the AGN and
non-AGN galaxies with increasing $T$ (see Fig.~\ref{f6}).  Consequently,
the spiral arm structure and large scale dust lane morphology (these two
factors contribute most of the large scale structure that we observe) are
similar in both groups of galaxies.

5)However, the $\sigma^{2}_{S100pc}$ values show a large, discontinuous
increase between the non-AGN late and early-type of galaxies, while they
remain roughly constant for active galaxies, both of early and late-type.

Consistent with previous studies (e.g. Malkan \etal 1998, Regan \&
Mulchaey 1999, Martini \& Pogge 1999), we find that all AGN show evidence
for significant central structure, irrespective of the host galaxy type.
The deviations are caused by localised regions of excess emission or
deficits, probably caused by dust absorption. The difference in the
morphology of the central region between AGN and non-AGN galaxies is
particularly strong in the case of early-type galaxies. The ``structure''
image of {\it all} the early-type, non-AGN galaxies (except from NGC 4649,
see below) looks like the structure image of NGC 4612 in Fig.~\ref{f3} (an
early-type galaxy itself). No significant deviations, of any kind, from
the fitted ellipses appears in the images of the other normal, early-type
galaxies. On the contrary, the structure images of {\it all} the
early-type, AGN galaxies show significant structures in their central
regions. An example is shown in Fig.~\ref{f2} (NGC 5273 is an early-type,
AGN galaxy). The central region (i.e. the innermost $100$ pc)
shows significant deviations from the isophotes, with many bright and dark
regions appearing, in a rather ``chaotic'' pattern. Large-scale dark
``lanes'', which extend up to the innermost region are also evident.  The
only exception among the early-type, non-AGN galaxies is NGC~4694, an H II
galaxy according to Ho \etal (1997a), which shows large amplitude
structure in its central regions (in fact, its $\sigma^{2}_{S100pc}$ value
is the largest among the early-type galaxies), similar to the structure
that is seen in the central region of AGNs. We suspect that this galaxy
may have a misclassified Hubble type (indeed it has a much more peculiar
shape than the typical $T=-2$ galaxies) or a misclassified nuclear
activity type.

Due to the small size of the early type, AGN and non-AGN galaxy
samples, the significant differences that we find in the morphology of
their central regions should be considered with caution. For example,
although we find no statistically significant difference in the
distribution of host type for the two samples, approximately $44\%$ of the
early-type non-AGN are ellipticals ($T=-5$), while only $11\%$ of the
early-type AGN are ellipticals. This could explain in part the observed
difference in the distributions of $\sigma^{2}_{S}$ values. However, only
one of the remaining nine non-AGN galaxies (with $T>-5$ shows central
micro-structure (NGC~4694), while all eight early-type, AGN galaxies with
$T>-5$ show significant structure in their central regions. If confirmed
with the use of larger samples, this result is consistent with the
hypothesis that the presence of an active nucleus in early-type galaxies
is associated with the presence of material in them. It is possible that
all early-type galaxies host a supermassive black hole, but only in those
cases where a sufficient quantity of interstellar material has managed to
reach the innermost region (and fuel the central engine), an active
nucleus is exhibited.

We find a different picture in the case of late-type galaxies. Almost all
of them show large amplitude structure in their central regions.  In fact,
the average amplitude of the deviations in the 100 pc innermost region of
the late-type non-AGN galaxies is {\it larger} than the amplitude in the
AGN of the same morphological type (see Fig.~\ref{f6}). This is a puzzling
result, which implies that the presence of significant structure, and
hence of material in the inner region of these galaxies, does not result
in the presence of an active nucleus in them.

One possibility is that the mass of the black hole in late-type normal
galaxies is small and hence the luminosity of the nucleus is not large
enough to be detected. We investigated this possibility by computing the
mass of the putative black hole in the center of the galaxies according to
the formula

\begin{equation}
M_{BH} = 0.78 \times 10^8 M_{\sun} (\frac{L_{bulge}}{10^{10}
L_{\sun}})^{1.08}
\end{equation}

given in Kormendy \& Gebhardt (2001). This relation between the black hole
mass ($M_{BH}$) and the $B$ band luminosity of the bulge component of the
galaxy ($L_{bulge}$) is established the last few years from stellar,
ionized gas, and maser dynamics observations of many, mainly inactive or
weakly active galaxies. Using the values of the absolute bulge magnitude
($M_{bulge}$) for our sample of galaxies (taken from Ho \etal 1997a) we
calculated the luminosity of the bulge ($L_{bulge} = 10^{0.4(4.79 -
M_{bulge})}$)  and thus the black hole mass $M_{BH}$ for each galaxy. We
found that the black hole mass for the AGNs in our sample ranges from $7.4
\times 10^6 M_{\sun}$ to $6.8 \times 10^8 M_{\sun}$ with a median value of
$2.0 \times 10^8 M_{\sun}$. For the non-AGNs, assuming that they also host
an ``invisible'' black hole, the mass range would be from $1.0 \times 10^7
M_{\sun}$ to $2.2 \times 10^9 M_{\sun}$ with a median value of $8.2 \times
10^7 M_{\sun}$. Hence, on average, the active galaxies host a black hole
with a mass which is $\sim$ 2.5 times larger than the mass of the black
hole in the non-AGN galaxies in our sample. However, there is also a
considerable overlap between the black hole mass values that we compute
for the AGN and non-AGNs. Therefore, we conclude that most of the
late-type, non-AGN galaxies in our sample either do {\it not} host a
supermassive black hole, or, for some reason, although there is enough
material in the central region (i.e. within the innermost 100 pc), it
cannot fuel the central engine.

In order to investigate possible reasons that could prevent the
fueling of the central engine in the galaxies that show the largest
amplitude structure in their innermost 100 pc region, we compared the
morphology of the circumnuclear structures that we observe in the
late-type, AGN and non-AGN galaxies, in trying to find whether there exist
any systematic differencies. We could not identify any clear patterns that
appear exclusively in one of the two groups of galaxies. There are
galaxies in both groups which show nuclear dust spiral formations, which
some times appear to connect to larger scale dust lanes. In other cases,
irrespective of the galaxy's activity type, the distribution of the
structures follows a chaotic pattern, with no clear, overall formation.

Therefore, the only significant difference that appears to exist between
late-type, AGN and non-AGNs is the amplitude of the nuclear structures.
Perhaps there exists an active nucleus in the late-type, non-AGN in
our sample but, if the larger amplitude structure that we find in these
galaxies implies the existence of a larger amount of gas in their central
region, then this material, apart from fueling the central black hole, may
also obscure the central active nucleus (including the Narrow Line Region)
from our sight. At the same time, the large amounts of gas could result in
the formation of a large number of star forming regions which make these
galaxies look like H II galaxies. As the central gas content decreases,
the active nucleus will be revealed and, at the same time, the central
black hole will have increased its mass, rendering the galaxy a ``normal"
AGN.

\section{Summary}

Using archival {\it Hubble Space Telescope} images, we examined the
central morphology of 58 galaxies (23 AGN and 35 non-AGN). Using the
``ellipse fitting" technique, we ``uncovered'' hidden structures in the
innermost parts of the galaxies. In order to compare, in a quantitative
way, the structure seen in the samples of AGN and normal galaxies we
calculated their variance (a quantity that is proportional to their
amplitude normalised to the underlying galactic emission) and compared its
distribution for different subgroups of the galaxies. We found that {\it
all} AGNs show significant structure in their central 100 pc region. The
amplitude of the structures is more or less independent of their Hubble
type. When grouping the galaxies according to their Hubble type we found
that, contrary to early-type AGNs, early-type non-AGN galaxies show no
structure at all. This result is consistent with the hypothesis that all
early-type galaxies host a supermassive black-hole, but in only those
cases where there is significant amount of material in their central
regions they host an active nucleus. On the other hand, late-type galaxies
show significant nuclear structures irrespective of whether they are AGNs
or not. This implies that the presence of material in the inner region of
these galaxies does not result in the presence of an active nucleus as
well. Either not all late-type galaxies host a central black hole,
or, for some reason, contrary to what happens in AGN, the significant
amount of material on the scales of tens-of-parsecs does not make it down
to the scales of the central black hole. Another possibility is that the
large amount of gas and dust in them obscure the nucleus from our sight.

Our results are based on the use of small size samples, because the
the number of current HST, WFPC2 images of nearby galaxies that
satisfy the criteria listed in Section 2 is small.  Obviously, larger
samples are needed in order to confirm our results. We plan to repeat the
analysis that we presented in this work in the future when a larger number
of observations of AGN and non-AGN galaxies will be available, and
investigate in greater detail the differencies/similarities in the
morphology of the central region in these galaxies.

\begin{acknowledgements}
We wish to thank the referee (J. Mulchaey) for useful comments
and suggestions on the improvement of this paper. This paper has
also benefited a lot from discussions with N. Kylafis, J. 
Papamastorakis, V. Charmandaris and K. Xilouris.
\end{acknowledgements}

\end{document}